\documentclass[twocolumn,showpacs,preprintnumbers,amsmath,amssymb,prb,aps]{revtex4}

\usepackage{epsfig}
\usepackage{graphicx}
\usepackage{dcolumn}
\usepackage{bm}

\begin{document}

\title{Electronic structure of BaIrO$_3$: A first principle study using
local spin-density approximations}

\author{Kalobaran Maiti}

\affiliation{Department of Condensed Matter Physics and Material
Sciences, Tata Institute of Fundamental Research, Homi Bhabha
Road, Colaba, Mumbai - 400 005, INDIA}

\date{\today}

\begin{abstract}

We investigate the electronic structure of BaIrO$_3$, an
interesting compound exhibiting charge density wave transition in
its insulating phase and ferromagnetic transition at the same
temperature, using full potential linearized augmented plane wave
method within the local spin density approximations. The
ferromagnetic ground state could exactly be described in these
calculations and the calculated spin magnetic moment is found to
be small as observed in the magnetic measurements. Interestingly,
no signature of exchange splitting is observed in the density of
states corresponding to Ir 5$d$ and/or any other electronic
states. The small spin moment appears essentially due to unequal
population of the up- and down-spin Ir 5$d$ bands. Comparison of
the valence band density of states with the experimental spectral
functions suggests that a rigid shift of the Fermi level towards
higher energies in the calculated density of states provides a
good description of the experimental spectra. This indicates that
the intrinsic oxygen non-stoichiometry leads to electron doping in
the system and plays the primary role in determining the
electronic structure rather than the electron correlation effects
as often observed in other systems. The calculated results for Ba
5$p$ core levels show that the Madelung potential of one of the
three non-equivalent Ba atoms is different from that of other two
as predicted in the recent experiments.
\end{abstract}


\maketitle

\section{Introduction}

The compound, BaIrO$_3$ has drawn significant attention in the
recent time due to the observation of charge density wave (CDW)
transition despite insulating transport observed above and below
the CDW transition temperature ($T_c \approx$~185~K).\cite{cao1}
Interestingly, it also exhibits a ferromagnetic transition at the
same temperature (185~K) where CDW ground state sets in. The
saturation magnetic moment in the ferromagnetic phase, however, is
found to be very small ($\sim$0.03~$\mu_B$) compared to that
expected from the $t_{2g}^5$ electronic configuration
corresponding to Ir$^{4+}$.

\begin{figure}
\vspace{-4ex}
 \centerline{\epsfysize=4.0in \epsffile{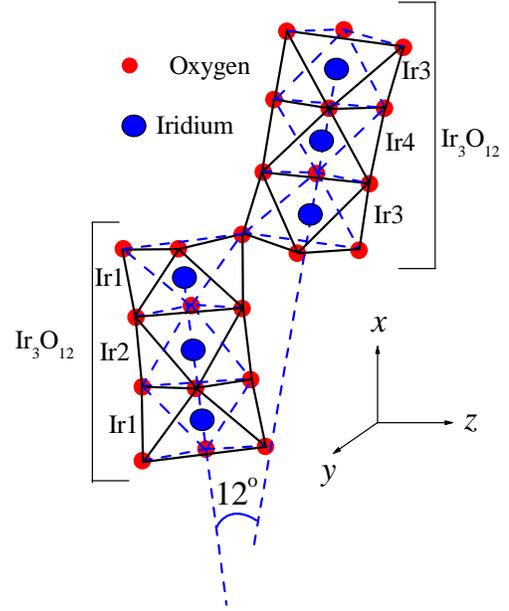}}
\vspace{-6ex}
 \caption{(color online) Schematic diagram of the Ir$_3$O$_{12}$
trimers in the monoclinic structure of BaIrO$_3$ (space group
B2/m). The trimers are connected by corner sharing and their long
axis are tilted by 12$^o$.}
 \vspace{-4ex}
\end{figure}

The crystal structure of BaIrO$_3$ is monoclinic (space group
B2/m; unique axis $c$). In this structure, IrO$_6$ octahedra form
Ir$_3$O$_{12}$ trimers by face sharing as shown in Fig.~1. The
Ir$_3$O$_{12}$ trimers are inter-linked by corner-sharing to form
columns parallel to $a$-axis. The axis of the connecting trimers
are marginally tilted ($\sim$12$^\circ$) as shown in the
figure.\cite{lindsay} This leads to the monoclinic structure with
the crystallographic angle, $\gamma$ = 103.27$^\circ$. This
typical structure makes it a quasi-one dimensional system as has
clearly been manifested in the anisotropic transport, magnetic and
optical properties.\cite{cao1} The structure becomes more complex
due to the twisting and buckling of the trimers resulting in the
multiplicity of Ir-O and Ba-O bond distances, thereby creating 4
types of Ir and 3 types of Ba sites.\cite{powell,siegrist}
Different non-equivalent Ir atoms are represented by Ir1, Ir2, Ir3
and Ir4 in the figure. Interestingly, each trimer consists of two
non-equivalent Ir atoms. For example, if Ir2 forms the central
octahedron of a trimer, it is connected by face sharing to the
other two octahedra formed by Ir1. Similarly, the neighboring
Ir$_3$O$_{12}$ trimer is formed by Ir3 and Ir4, where Ir4 forms
the central octahedron and Ir3 forms the other two octahedra.
Thus, the inter-trimer link is established along $a$ direction via
(Ir1)-O-(Ir3) connectivity.

Despite very short intra-trimer Ir-Ir distances compared to
Ir-metal and Ir-O-Ir angle close to 180$^\circ$ for inter-trimer
connections, BaIrO$_3$ exhibits insulating transport behavior. In
addition, there are multiple metal-insulator (MI) transitions
within the CDW ground state. While the occurrence of CDW
transitions in the insulating phase is surprising, a recent
high-resolution photoemission study\cite{bairprl} reveals the
signature of finite density of states at the Fermi level,
$\epsilon_F$, at room temperature and it was shown that the CDW
ground state evolves by opening a soft gap at $\epsilon_F$ across
CDW transition temperature. Interestingly, the intensity at
$\epsilon_F$ in the photoemission spectra at room temperature ($T
> T_C$) is significantly small consistent with the
observation of very small electronic specific heat
coefficient\cite{cao1} ($\sim$~1~mJ/mole.K$^2$). In contrast,
previous theoretical study using extended H\"{u}ckel tight-binding
electronic band structure calculations\cite{bairband} exhibits
large intensity at $\epsilon_F$ as expected for a $t_{2g}^5$
electronic configuration corresponding to Ir$^{4+}$. In addition,
the analysis of the core level spectra in the photoemission
measurements\cite{bairprl} predicts the existence of two different
kinds of Ba sites at room temperature, which become similar before
the onset of CDW ground state. All these
anomalies\cite{cao1,lindsay,powell,siegrist,bairprl,bairband,chamberland,cao2}
in the transport, magnetic and spectroscopic data reveal that
BaIrO$_3$ is truly an exotic material.

Since, the 5$d$ orbitals in 5$d$ transition metal oxides are
highly extended compared to that in 3$d$ or 4$d$ systems, it is
expected that the electron correlation strength will be
significantly weak and that the {\em ab initio} approaches will be
successful to describe the electronic properties in this system.
In this study, we therefore, investigate the electronic structure
of BaIrO$_3$ using {\em state-of-the-art ab initio} approaches.
The calculations for various magnetic configurations suggest that
the ground state of BaIrO$_3$ is ferromagnetic exhibiting small
magnetic moments consistent with the experimental
observations.\cite{cao1} The analysis of the valence band density
of states in comparison with the experimental photoemission
results\cite{bairprl} suggests that the small intensity at
$\epsilon_F$ in the photoemission spectra and the insulating
transport behavior are presumably related to the pinning of
$\epsilon_F$ at the upper edge of $t_{2g}$ band due to the
intrinsic non-stoichiometry of the samples\cite{powell} and these
edge states are localized due to disorder/crystallographic
distortions.\cite{mott,anderson} The calculations for the Ba 4$p$
core levels exhibit signature of two non-equivalent Ba sites with
respect to the Madelung potential at the Ba sites in the crystal
structure at room temperature.

\section{Theoretical methods}

The electronic band structure calculations were carried out using
full potential linearized augmented plane wave (FPLAPW) method
(Wien2k software\cite{wien}) within the local spin density
approximations (LSDA). The crystal structure considered for these
calculations has B2/m space group with the $c$-axis as the unique
axis. The lattice parameters ($a$ = 10.005 \AA, $b$ = 15.174 \AA,
$c$ = 5.571 \AA, $\alpha$ = $\beta$ = 90$^\circ$ and $\gamma$ =
103.27$^\circ$) and the atomic positions used for all the
calculations correspond to the room temperature structure reported
for well characterized polycrystalline and single crystalline
samples in References [1,3]. It is to note here that most of these
studies for the crystal structure analysis report the space group
as C2/m with $b$ axis as the unique axis, which has been
transformed to B2/m space group by coordinate transformations
($a^\prime b^\prime c^\prime \Leftrightarrow acb$) in order to
satisfy the requirement of the Wien2k software. The muffin-tin
radii ($R_{MT}$) for Ba, Ir and O were set to 1.217 \AA, 1.111
\AA\ and 0.847 \AA, respectively. The convergence for different
calculations were achieved considering 512 $k$ points within the
Brillouin zone. The error bar for the energy convergence was set
to $<$~1~meV.

\begin{figure}
\vspace{-6ex}
 \centerline{\epsfysize=5.0in \epsffile{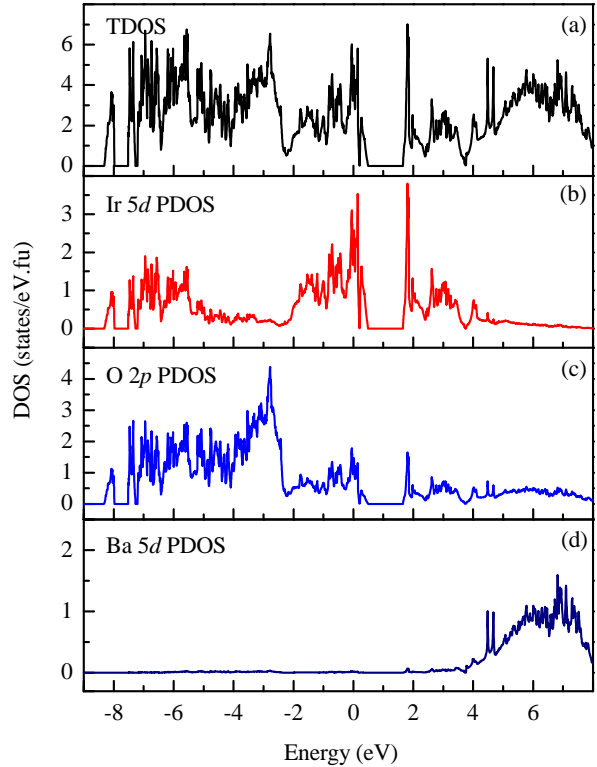}}
\vspace{-8ex}
 \caption{(color online) Calculated (a) TDOS, (b) Ir 5$d$ PDOS, (c) O
2$p$ PDOS and (d) Ba 5$d$ PDOS for the non-magnetic ground state.
The large intensity at the Fermi level suggests a metallic ground
state in contrast to the experimental observations. The states at
the Fermi level have primarily Ir 5$d$ character with O 2$p$
states appearing below -2.3~eV.}
 \vspace{-4ex}
\end{figure}

\section{Results and Discussions}

In Fig.~2, we show the density of states calculated for the
non-magnetic ground state using the lattice parameters
corresponding to room temperature structure as mentioned in the
previous section. The total density of states (TDOS), Ir 5$d$
partial density of states (PDOS), O 2$p$ PDOS and Ba 5$d$ PDOS are
plotted in Fig.~2(a), 2(b), 2(c) and 2(d), respectively. Ba 5$d$
electronic states contribute in the energy range 4~eV above the
Fermi level as shown in Fig.~2(d) with almost no contribution at
lower energies. O 2$p$ contribution in the Ba 5$d$ dominated
energy region is very little, which is expected for the 5$d$
energy level in this heavy alkaline earth atom compared to the
cases observed for lighter alkaline earth atoms such as Sr in
Sr-compounds or Ca in Ca-compounds having unoccupied 4$d$ and 3$d$
levels, respectively.\cite{andersen} This suggests that the role
of Ba-O covalency in the electronic/crystal structure is not
significant in this system.

\begin{figure}
\vspace{-6ex}
 \centerline{\epsfysize=4.5in \epsffile{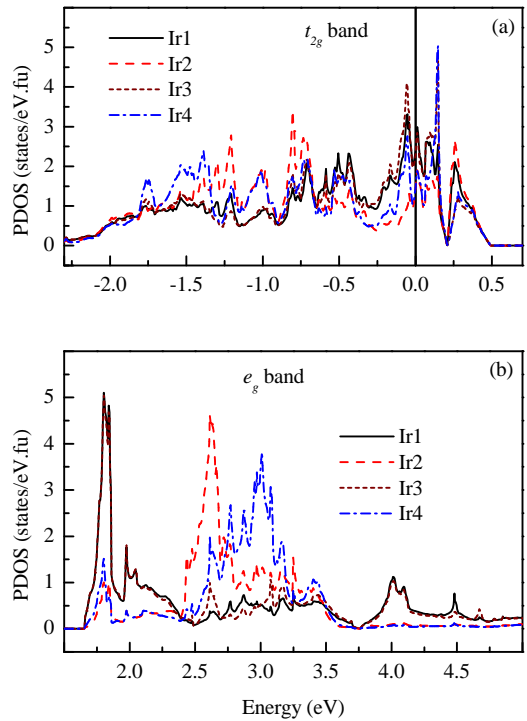}}
\vspace{-6ex}
 \caption{(color online) Ir 5$d$ band with (a) $t_{2g}$ symmetry and
(b) $e_g$ symmetry for all the 4 non-equivalent Ir atoms in the
structure. The assignment of Ir atoms are made as shown in
Fig.~1.}
 \vspace{-4ex}
\end{figure}

There are three groups of features observed in TDOS in the
occupied part of the electronic structure shown below $\epsilon_F$
(=~0~eV). The energy range between -5.5~eV to -2.3~eV is primarily
dominated by the O 2$p$ PDOS with negligible contributions from Ir
5$d$ PDOS. This suggests that the electronic states in this energy
range are not bonded with the Ir 5$d$ states and are known as
non-bonding O 2$p$ states. In the energies below -5.5~eV
(-5.5~eV$>\epsilon>$-8.5~eV), O 2$p$ states have the largest
intensity with finite contributions from Ir 5$d$ electronic
states. Thus, these features can be assigned as bonding bands
having primarily O 2$p$ character. The anti-bonding bands having
$t_{2g}$ symmetry appears in the energy range -2.5~eV to +0.5~eV
(total width $\sim$~3~eV). The $e_g$ band appears above the Fermi
level with a minimum excitation gap of about 1.2~eV from the upper
edge of $t_{2g}$ band. The DOS in these regions are essentially
contributed by the Ir 5$d$ electronic states with minimal
contributions from O 2$p$ states. Thus, the electronic states at
and around $\epsilon_F$ have primarily Ir 5$d$ character, which
naturally determines the transport, magnetic and other
thermodynamic properties. Interestingly, DOS at $\epsilon_F$ is
found to be significantly large as also observed
before\cite{bairband} suggesting highly metallic behavior of the
system in contrast to the insulating transport, small specific
heat coefficient observed\cite{cao1} and the high-resolution
photoemission results recently published.\cite{bairprl}

In order to investigate the contribution from various
non-equivalent Ir sites separaterly, we plot the $t_{2g}$ and
$e_g$ band in Fig.~3(a) and 3(b), respectively. The total $t_{2g}$
bandwidth for each of the Ir atoms is found to be close to 3~eV
spanning the same energy region. The PDOS corresponding to Ir2 and
Ir4 are almost uniformly distributed over the whole energy range.
However, Ir1 and Ir3 exhibit largest intensity close to the Fermi
level suggesting largest contribution to the conduction electrons.
Such a scenario is expected since inter Ir$_3$O$_{12}$ trimer
connectivity is established via (Ir1)-O-(Ir3) superexchange
interaction as shown in Fig.~1 and has largest extended character.
Ir2 and Ir4 provide intra-trimer connectivity via $d-d$
interactions with Ir1 and Ir3, respectively. Thus, there is a
relative shift between the center of gravity of Ir1, Ir3
contributions and Ir2, Ir4 contributions. The features in the
energy range $\epsilon
<$~-0.75~eV has largest contribution from Ir2 and Ir4 sites and
for $\epsilon >$~-0.75~eV, the intensity arises primarily from Ir1
and Ir3 contributions.

\begin{figure}
\vspace{-6ex}
 \centerline{\epsfysize=5.0in \epsffile{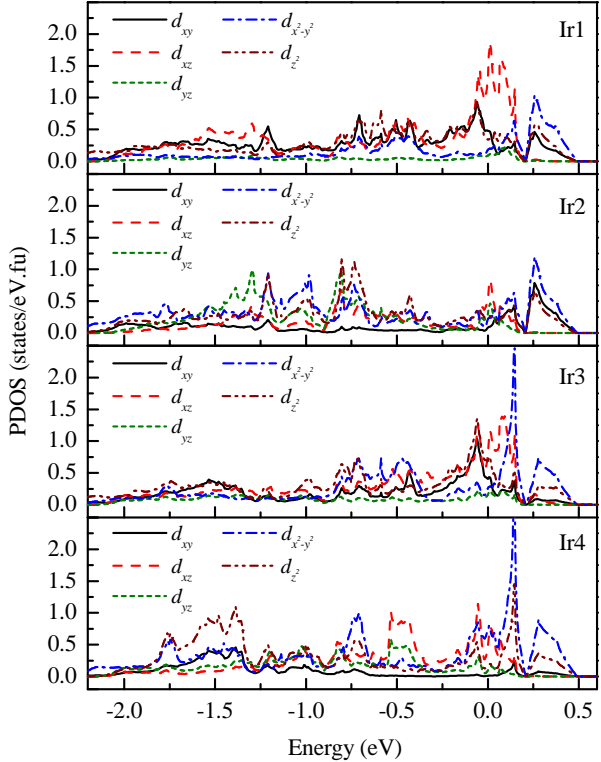}}
\vspace{-9ex}
 \caption{(color online) PDOS for the $t_{2g}$ band in the
non-magnetic ground state having $d_{xy}$ character (solid line),
$d_{xz}$-character (dashed line), $d_{yz}$-character (short dashed
line), $d_{x^2-y^2}$-character (dot-dashed line) and
$d_{z^2}$-character (dot-dot-dashed line) are shown separately for
four non-equivalent Ir atoms.}
 \vspace{-4ex}
\end{figure}

The effective width of the $e_g$ bands is significantly narrower
in this system as shown in Fig.~3(b). While Ir2 and Ir4 are
primarily contributing in the energies 2.5~eV - 3.5~eV above
$\epsilon_F$, the Ir1 and Ir3 contributions appear below 2.5~eV.
Interestingly, Ir1 and Ir3 are found to be very similar in both
the energy regions, and Ir2 and Ir4 have similar DOS
distributions.

Since the axis of any of the IrO$_6$ octahedra are not aligned
along any of the crystal axis as shown in Fig.~1, the $t_{2g}$
bands will have contributions from all the $d$-orbital states
($d_{xy}$, $d_{yz}$, $d_{xz}$, $d_{x^2-y^2}$ and $d_{z^2}$)
defined in the axis system shown in Fig.~1. The projected partial
density of states corresponding to these orbital are shown
separately for each of the Ir sites in Fig.~4. Interestingly, the
primary contribution ($\sim$~35.77\%) at $\epsilon_F$ comes from
$d_{xz}$ electronic states located at Ir1 and Ir3 sites. $d_{xy}$
and $d_{z^2}$ at Ir1 and Ir3 sites also exhibit large
contributions (14.3\% and 16.6\% respectively) at $\epsilon_F$.
The largest contributions from Ir2 and Ir4, however, arise from
$d_{xz}$ (8.4\%) and $d_{x^2-y^2}$ (7.6\%) electronic states.
These numbers clearly manifest that the electronic states
containing the quasi-one dimensional axis of the crystal structure
($x$-axis) have largest contribution at $\epsilon_F$ and hence
clearly manifests the signature of large anisotropy observed in
various bulk properties.\cite{cao1}

\begin{figure}
\vspace{-6ex}
 \centerline{\epsfysize=5.0in \epsffile{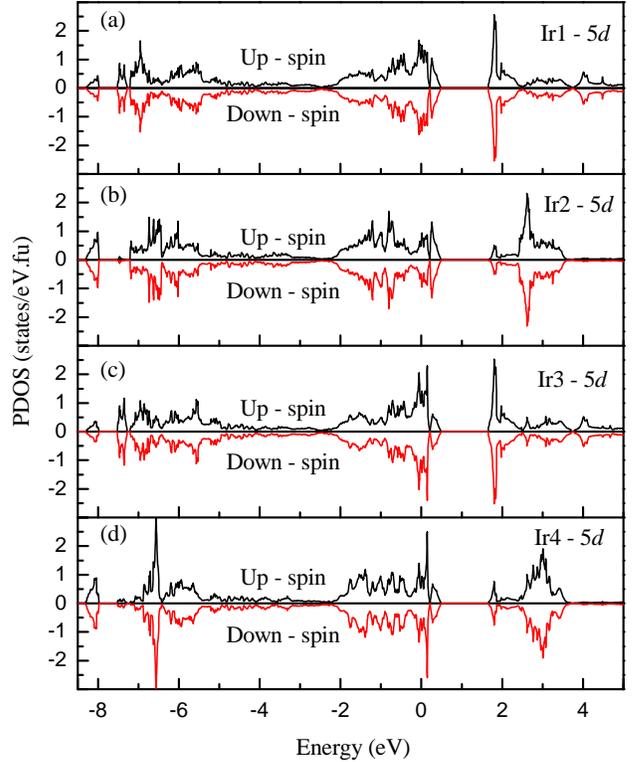}}
\vspace{-9ex}
 \caption{(color online) Ir 5$d$ PDOS corresponding to ferromagnetic
ground state. Up- and down-spin contributions are shown by
positive and negative y-axis, respectively. None of the cases show
signature of exchange splitting.}
 \vspace{-4ex}
\end{figure}

We now turn to the understanding of the magnetic behavior of
BaIrO$_3$. We carried out the spin-polarized calculations using
the same crystal structure as in the case of non-magnetic
calculations and use local spin density approximations. The
magnetic moment has been defined by the difference in occupancy of
the majority (up) and minority (down) spin states. Spin-orbit
interaction was not considered in these calculations.
Interestingly, the calculations for various magnetic
configurations reveal that the ferromagnetic ground state is
3.25~meV lower than (error bar used for convergence is $<$~1~meV)
the total energy found for the non-magnetic ground state. This
energy difference is about 20\% of $k_BT_C$ ($T_C$ = 185~K). This
calculation, thus, once again establish that LSDA approximations
are quite accurate in capturing the magnetic ground state of
various transition metal oxides as has also been observed in 3$d$
transition metal oxides\cite{dds3d} despite the fact that the
electron correlation effects are underestimated in these
calculations. The total spin magnetic moment is found to be about
1.34$\times$10$^{-3}$~$\mu_B$. This is significantly small
compared to that expected for $t_{2g}^5$ electronic configuration
for Ir$^{4+}$. Such a small value of the spin moment appears
presumably due to the highly extended nature of the 5$d$ orbitals
leading to large kinetic energy of the associated electrons. The
calculated spin moment is consistent with the experimental
observation of small saturation magnetic moment
($\sim$~0.03~$\mu_B$) in the magnetic measurements of single
crystalline BaIrO$_3$; this moment could be measured only after
applying a magnetization field as high as 20~T.\cite{cao1} No
magnetic contribution is found from the Ba-related bands, as all
these bands are either completely occupied and appear much below
$\epsilon_F$ or completely unoccupied appearing far above
$\epsilon_F$. The contribution from O 2$p$ states is of the order
of 10$^{-4} \mu_B$. The primary contribution to the total magnetic
moment arises from the spin polarization of Ir 5$d$ partial
density of states. Interestingly, all the Ir atoms are found to
order ferromagnetically and the magnetic moment at the central Ir
site of the Ir$_3$O$_{12}$ trimer is almost the half of that
observed at the other Ir sites.

\begin{figure}
\vspace{-6ex}
 \centerline{\epsfysize=4.5in \epsffile{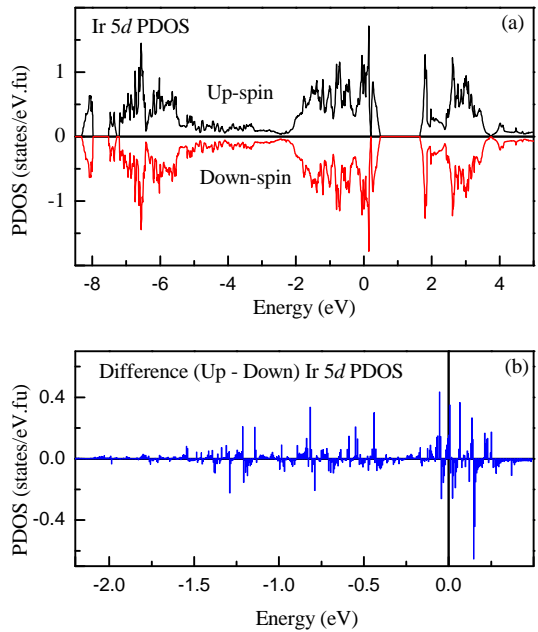}}
\vspace{-14ex}
 \caption{(color online) Total Ir 5$d$ PDOS corresponding to
ferromagnetic ground state. The difference between up- and
down-spin PDOS are shown in (b). The spin moment appears due to
unequal occupancy of the up- and down-spin bands as shown in the
figure.}
 \vspace{-2ex}
\end{figure}

In Fig.~5, we show the spin-polarized Ir 5$d$ partial density of
states for the ferromagnetic ground state, where 5(a), 5(b), 5(c)
and 5(d) show the results corresponding to Ir1, Ir2, Ir3 and Ir4
respectively. The down-spin contributions are shown by inverting
the $y$-axis in the same figure with same energy scale for up-spin
contributions shown along $x$-axis. No signature of exchange
splitting is observed at any of the Ir sites. The up- and
down-spin DOS reveal almost identical structures in all the cases.
In order to investigate the origin of the spin moment, we show the
total Ir 5$d$ PDOS for both up- and down-spin states in Fig.~6(a)
and the difference between up- and down-spin contributions in
Fig.~6(b). It is clear that the observed magnetic moment appears
due to small asymmetry of the up- and down-spin density of states
as a function of energy. Thus, the occupancy of the up- and
down-spin contributions is different leading to a small
spin-moment.

Now, we compare the calculated results with the experimental
photoemission spectra published recently for
BaIrO$_3$.\cite{bairprl} In Fig.~7(a), we show the Ir 5$d$ band
obtained from He~{\scriptsize II} spectra\cite{bairprl} and
overlap with the calculated $t_{2g}$ band of Ir 5$d$ PDOS. While
the calculated results exhibit large intensity at the Fermi level
suggesting highly metallic behavior of the system, the
experimental Ir 5$d$ feature at room temperature exhibit only very
small intensity. The total width of the occupied part in the
calculated PDOS is also substantially smaller compared to the Ir
5$d$ signal in the experimental spectra. Such observations of
significant reduction of spectral weight at the Fermi level and
the population of the higher binding energy region often
attributed to the underestimation of electron-electron Coulomb
repulsion effect in the {\it ab initio} approaches. However, such
a strong correlation effect in these highly extended 5$d$
electrons is unexpected.

\begin{figure}
\vspace{-10ex}
 \centerline{\epsfysize=4.5in \epsffile{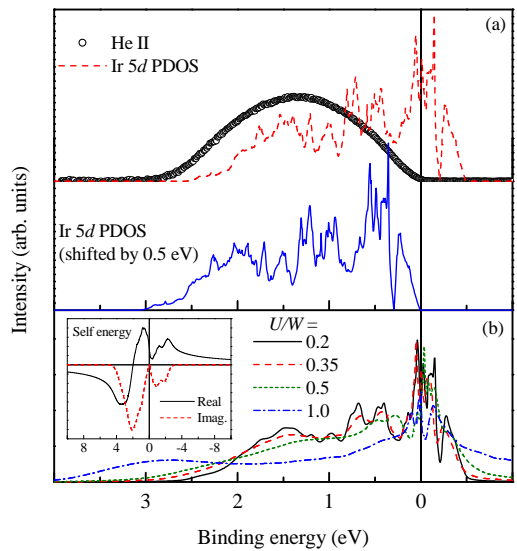}}
\vspace{-20ex}
 \caption{(color online) (a) The $t_{2g}$ band (dashed line) obtained
from {\it ab initio} calculations is compared with the
experimental Ir 5$d$ band (circles) delineated from the
He~{\scriptsize II} spectrum. The experimental results are adopted
from Ref. [5]. The calculated DOS is significantly different from
the experimental ones. The calculated DOS shifted by 0.5~eV
towards higher binding energies (solid line) reveal better
description of the experimental spectrum. (b) Spectral functions
obtained by introducing electron correlation effects into the self
energy by second order perturbative approach using {\it ab initio}
results. The change in $U/W$ leads to a spectral weight transfer
toward higher binding energies. The inset shows the real (solid
line) and imaginary (dashed line) part of the self energy.}
 \vspace{-2ex}
\end{figure}

It is to note here that the difference between experimental bulk
spectral functions and the {\em ab initio} results could well be
explained in the 4$d$ transition metal oxides\cite{prbruth} using
perturbative approach to introduce the correlation effect in the
self energy of the system. The effective correlation strength,
$U/W$ ($U$ = electron-electron Coulomb repulsion strength, $W$ =
bandwidth) was found to be small ($U/W \sim$~0.2) in these
systems. We thus calculate the spectral function for BaIrO$_3$ in
the same way, using the second order perturbation method employed
by Treglia {\em et al.}\cite{treglia} In this method, the spectral
function can be expressed as, $$f(\epsilon) = -{1\over \pi} Im
\sum_k G_k(\epsilon)$$ where, $G_k(\epsilon)$ is the retarded
Green's function representing the many electron system and is
given by
$$G_k(\epsilon)={{1}\over{(\epsilon-\Sigma_k{(\epsilon)}-\epsilon_k)}}$$
where, $\Sigma_k{(\epsilon)}$ is the self energy of the system.
For small $U$, $\Sigma_k{(\epsilon)}$ can be calculated using
perturbation approach upto the second order term\cite{treglia,dds}
within the local approximations. We have used Ir $t_{2g}$ PDOS
shown in Fig.~7(a) as the band DOS for these calculations. The
calculated spectral functions for different $U/W$ are shown in
Fig.~7(b) and the calculated self energy is shown in the inset of
the same figure. The increase in $U/W$ leads to a significant
spectral weight transfer to higher energies. The new features
appearing at higher binding energies due to such spectral weight
transfer is known as incoherent feature and represent the
electronic states essentially localized due to such correlation
effect. The feature at the Fermi level represent the delocalized
states and is known as coherent feature. The total weight of the
coherent feature diminishes gradually with the increase in $U/W$.
However, the intensity at the Fermi level remains almost the same
for $U/W$ as large as 1.0. The spectral function obtained in this
process could not generate the spectral lineshape observed in the
experimental spectrum for any value of $U/W$. This suggests that
this procedure may not be adequate to determine the experimental
results.

Interestingly, a rigid shift of the calculated $t_{2g}$ band by
0.5~eV towards higher binding energies provide a significantly
good description of the experimental spectrum. The width and
spectral distribution in the shifted spectral function shown in
Fig.~7(a) is remarkably similar to that observed in the Ir 5$d$
band in the He~{\scriptsize II} spectrum. In order to investigate
this further we compare the O 2$p$ PDOS with the experimental
spectra as it is well known that the {\it ab initio} calculations
are quite successful to provide a good description of the O 2$p$
band even in 3$d$ transition metal oxides.\cite{dds3d} We show the
experimental Al~$K\alpha$ and He~{\scriptsize I} spectra in Fig.~8
adopted from Ref. [5]. The changes in these experimental spectra
in the energy range between 3~eV to 9~eV binding energies
concomitant to the change in relative photoemission cross
section\cite{yeh} of the O 2$p$ and Ir 5$d$ electronic states
suggest that the features in 3-6~eV binding energies appear
primarily due to the excitations of O 2$p$ non-bonding states and
the intensity at higher binding energies are due to the bonding
electronic states having large O 2$p$ character. Interestingly,
the calculated O 2$p$ PDOS shows the intense non-bonding O 2$p$
features around 2.3-5.5~eV binding energies with the peak at
2.8~eV binding energy (see Fig.~2), where the dip appears in the
He~{\scriptsize I} spectrum. This clearly suggests that it is
necessary to shift the O 2$p$ PDOS by at least 0.5~eV towards
higher binding energies to compare with the experimental spectra.
The shifted O 2$p$ PDOS shown by solid line in Fig.~8 provide a
good description of the experimental results.

\begin{figure}
\vspace{-6ex}
 \centerline{\epsfysize=4.5in \epsffile{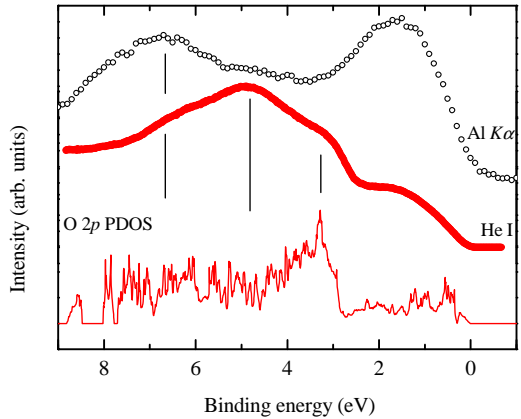}}
\vspace{-36ex}
 \caption{(color online) Experimental He~{\scriptsize I} (solid
circles) and Al~$K\alpha$ spectra (open circles) adopted from Ref.
[5] are compared with the calculated O 2$p$ PDOS (soild line)
shifted by 0.5~eV towards higher binding energies.}
 \vspace{-2ex}
\end{figure}

Thus, the results in Fig.~7 and Fig.~8 suggest that the Fermi
level is shifted to the upper edge of the $t_{2g}$ band presumably
due to the electron doping into the system. Such a shift would
stabilize the system due to the closeness to the fully filled
electronic configuration resulting to lowering in energy.
Interestingly, various experimental results indeed indicate that
the oxygen content is often found to be less than
3.0,\cite{powell,chamberland} which effectively leads to electron
doping into the Ir 5$d$ band. In addition, the transport
measurements always exhibit insulating temperature
dependence.\cite{cao1} Such a behavior is not unusual since it is
natural that the small intensity at the upper edge of the $t_{2g}$
band will be localized due to the disorder and/or crystallographic
distortions\cite{mott,anderson} in this quasi-one dimensional
structure leading to an insulating nature. All these observations,
thus, clearly indicate that the electron correlation may not be
the primary origin for the differences between experimental and
theoretical results. The electron doping due to the oxygen
non-stoichiometry intrinsic to BaIrO$_3$ possibly plays the key
role in determining the electronic structure and subsequently
various physical properties in this system.

\begin{figure}
\vspace{-9ex}
 \centerline{\epsfysize=4.4in \epsffile{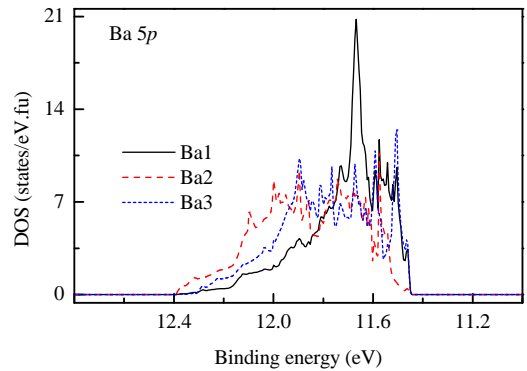}}
\vspace{-36ex}
 \caption{(color online) Calculated Ba 5$p$ DOS corresponding to 3
non-equivalent Ba atoms present in the monoclinic structure. The
results show that Ba2 contributions appear at higher binding
energies compared to that in other Ba atoms}
 \vspace{-2ex}
\end{figure}

We now turn to the question of multiple Ba core level signals in
the experimental spectra.\cite{bairprl} We show the DOS
corresponding to Ba 5$p$ core levels in Fig.~9. There are three
non-equivalent Ba sites in the structure with respect to the Ba-O
bond length and the symmetries. It is clear that the DOS
corresponding to Ba1 and Ba2 appear in the same energy range.
While the width of the DOS corresponding to Ba2 5$p$ states is
similar to those corresponding to Ba1 and Ba3, the peak position
appear at higher binding energies. The energy difference between
these spectral features is about 0.25~eV. This observation clearly
suggests that the Madelung potential at all the Ba sites are not
same and is consistent with the experimental observation of two
non-equivalent features corresponding to each Ba core level
photoemissions. Such relative shift of the DOS corresponding to
different non-equivalent Ir atoms is not observed in these
calculations (e.g. see Fig.~3 for the $t_{2g}$ bands corresponding
to 4 non-equivalent Ir sites).

\section{Conclusions}

In summary, we investigate the detailed electronic structure of
BaIrO$_3$ in this study using full potential linearized augmented
plane wave method within the local spin density approximations.
The calculations for various magnetic configurations reveal lowest
energy for the ferromagnetic ground state and is consistent with
the experimental observations. The ferromagnetic ground state
energy calculated using room temperature crystal structure is
found to be 3.25~meV lower than the non-magnetic ground state. No
exchange splitting is observed in any of the electronic states
involved in forming the valence band. The small magnetic moment
appears due to unequal population of the different Ir 5$d$ spin
density of states.

Comparison with the photoemission spectra suggests that electron-
electron Coulomb repulsion may not be the key factor in
determining the electronic structure as expected for the highly
extended nature of the 5$d$ electronic states. The discrepancy
between the experimental spectra and the calculated results
appears to arise primarily from the intrinsic non-stoichiometry of
BaIrO$_3$ leading to electron doping into the $t_{2g}$ band. The
calculations for Ba 5$p$ core levels exhibit the signature of
multiple non-equivalent Ba sites in the room temperature structure
as observed experimentally. Thus, these results provide a
significant advance towards understanding the electronic
properties of the exotic compound, BaIrO$_3$.


\begin{thebibliography}{}


\bibitem{cao1} G. Cao, J.E. Crow, R.P. Guertin, P.F. Henning, C.C. Homes,
M. Strongin, D.N. Basov, and E. Lochner, Solid State Commun. {\bf
113}, 657 (2000).

\bibitem{lindsay} R. Lindsay, W. Strange, B.L. Chamberland, and R O Moyer,
Jr., Solid State Commun. {\bf 86}, 759 (1993).

\bibitem{powell} A.V. Powell and P.D. Battle, J. Alloys and
Compounds {\bf 232}, 147 (1996).

\bibitem{siegrist} T. Siegrist and B.L. Chamberland, J.
Less-Common Met. {\bf 170}, 93 (1991).

\bibitem{bairprl} K. Maiti, R.S. Singh, V.R.R. Medicherla, S.
Rayaprol, and E.V. Sampathkumaran, Phys. Rev. Lett. {\bf 95},
016404 (2005).

\bibitem{bairband} M.-H. Whangbo and H.-J. Koo, Solid State
Commun.{\bf 118}, 491 (2001).

\bibitem{chamberland} B.L. Chamberland, J. Less-Common Met. {\bf
171}, 377 (1991).

\bibitem{cao2} G. Cao, X. N. Lin, S. Chikara, V. Durairaj, and E. Elhami,
Phys. Rev. B {\bf 69}, 174418 (2004); A. Ohnishi, M. Sasaki, Y.
Kuroda, M. Sato, Y. Isobe, and G. Cao, Physica B {\bf 329}, 930
(2003).

\bibitem{mott} N.F. Mott {\em Metal-Insulator Transitions}, 2nd
ed. (Taylor \& Francis, London, 1990).

\bibitem{anderson} P.W. Anderson, Phys. Rev {\bf 109}, 1492
(1958).

\bibitem{wien} P. Blaha, K. Schwarz, G.K.H. Madsen, D. Kvasnicka, and J. Luitz,
{\bf WIEN2k}, An Augmented Plane Wave + Local Orbitals Program for
Calculating Crystal Properties (Karlheinz Schwarz, Techn.
Universit\"{a}t Wien, Austria), 2001. ISBN 3-9501031-1-2.

\bibitem{andersen} E. Pavarini, S. Biermann, A. Poteryaev, A. I. Lichtenstein,
A. Georges, and O. K. Andersen, Phys. Rev. Lett. {\bf 92}, 176403
(2004).

\bibitem{dds3d} D. D. Sarma, N. Shanthi, S. R. Barman, N. Hamada, H. Sawada,
and K. Terakura, Phys. Rev. Lett. {\bf 75}, 1126 (1995).

\bibitem{prbruth} K. Maiti and R.S. Singh, Phys. Rev. B {\bf 71}, 161102(R) (2005).

\bibitem{treglia} G. Treglia, F. Ducastelle, and D.G. Spanjaard, J. Physique
{\bf 41}, 281 (1980); G. Treglia, F. Ducastelle, and D. Spanjaard,
Phys. Rev. B {\bf 21}, 3729 (1980).

\bibitem{dds} D. D. Sarma, F. U. Hillebrecht, W. Speier, N. M{\aa}rtensson,
and D. D. Koelling, Phys. Rev. Lett. {\bf 57}, 2215 (1986).

\bibitem{yeh} J.J. Yeh and I. Lindau, {\it At. Data and Nucl. Data Tables} {\bf
32}, 1 (1985).

\end{thebibliography}
\end{document}